\documentclass[useAMS,usenatbib]{mnras}
\usepackage{epsfig}
\usepackage{color} 
\usepackage{natbib}
\usepackage{booktabs,caption}
\usepackage[flushleft]{threeparttable}
\usepackage{graphicx}
\usepackage{amsmath,amssymb}
\usepackage{float,lscape}
\usepackage{longtable}
\usepackage[toc,page]{appendix}

\title[XRG]{Spectral index variation across X-shaped radio galaxies}

\author[Patra~et~al.]{
Dusmanta Patra$^{1}$,\thanks{E-mail: dusmanta.phy@gmail.com}
Ravi Joshi$^{2}$,\thanks{E-mail: rvjohshirv@gmail.com}
Gopal-Krishna$^{3}$
\\
$^{1}$S N Bose National Centre for Basic Sciences, Kolkata, \it{700106}; India\\
$^{2}$Indian Institute of Astrophysics, Koramangla, Bangalore,  \it{560034}; India\\
$^{3}$UM-DAE Centre for Excellence in Basic Sciences, Mumbai, \it{400098}; India
}

\begin{document}
\date{Accepted ---. Received ---; in original form ---}

\pagerange{\pageref{firstpage}--\pageref{lastpage}} \pubyear{2019}

\maketitle

\label{firstpage}
\begin{abstract}
The formation mechanism of the enigmatic subclass of radio galaxies, called `X-shaped radio galaxies' (XRGs), or `winged' radio galaxies, which account for $\sim 10\%$ of the radio galaxy population, can be effectively constrained using the radio spectral-index distribution across their twin pairs of radio lobes. If indeed, the existing claims of no systematic spectral index difference between the wing and the associated primary lobe are valid in general, this would provide impetus to the XRG model attributing their origin to an unresolved binary of active supermassive black holes within the nucleus of the host galaxy. To investigate this interesting possibility, we have mapped spatial variation of spectral index for a well-defined sample of 25 XRGs, by combining their 1.4 GHz VLA (FIRST survey)/uGMRT maps with their 144 MHz maps (LoTSS-DR2). This has yielded the best available combination of sensitivity, angular resolution, frequency range and sample size, for spectral mapping of an XRG sample. A rich diversity of spectral index patterns is thus revealed in our XRG sample, but we find at most one case where a secondary lobe (wing) exhibits a flatter spectrum compared to its associated primary lobe. We conclude that such a spectral pattern is exceedingly rare and by no means a common trait of XRGs.
\end{abstract}

\section{Introduction}
\label{sec:intro}
`X-shaped radio galaxies' (XRGs), also called `winged' radio galaxies, are a small but much discussed subclass of extragalactic radio sources. This is because, in addition to the pair of `active’ (primary) radio lobes, they also exhibit a pair of secondary radio lobes (`wings') of lower surface-brightness, oriented at a large angle to the primary lobe pair (\citealt{Leahy1984MNRAS.210..929L,Leahy1992ersf.meet..307L}; also, \citealt{Hogbom1974A&A....34..341H}).
In a subset of such sources, the secondary lobes start off near the outer extremities of the primary lobes, leading to a `Z-symmetric' morphology and such sources are termed  `Z-shaped' radio galaxies (ZRGs) (e.g. \citealt{Zier2005MNRAS.364..583Z,Saripalli2018ApJ...852...48S, Joshi2019ApJ...887..266J}; also, \citealt{Riley1972MNRAS.157..349R}).
XRGs comprise $\sim 10\%$ of the radio galaxy population \citep{Yang2019ApJS..245...17Y,Cheung2007AJ....133.2097C,Bera2020ApJS..251....9B}.
Since the wings in XRGs are always found to lack a terminal hot spot, the Fanaroff-Riley classification (\citet{Fanaroff1974MNRAS.167P..31F} of XRGs is based on their primary lobes.

Despite several models proposed, a consensus is yet to emerge about the formation mechanism of XRGs \citep[see,][for a review]{Gopal-Krishna2012RAA....12..127G}. Prominent among these are the `backflow diversion' model where the wings are produced as a result of diversion of the synchrotron plasma streams flowing back from the hot spots of the two active lobes, as they impinge upon the (asymmetric) gaseous halo (ISM) of the parent galaxy 
\citep{Leahy1984MNRAS.210..929L, Leahy1992ersf.meet..307L, Capetti2002A&A...394...39C, Cotton2020MNRAS.495.1271C}.
The post-diversion outward flow could be supported by buoyancy forces (e.g., 
\citep{Worrall1995ApJ...449...93W,Capetti2002A&A...394...39C,Saripalli2009ApJ...695..156S,Hodges2011ApJ...733...58H},
and/or by the magnetic field pressure in a kpc-scale gaseous disk associated with the host galaxy \citep{Gopal-Krishna-Dhabade_2022A&A...663L...8G}.
Another widely discussed model for XRGs envisions a rapid reorientation of the jet pair due to `spin-flip' of the central supermassive black hole (SMBH), probably due to its merger with another SMBH, and thereby giving rise to a new pair of active lobes while the older lobes decay and appear as a pair of wings \citep{Rottmann2001PhDT.......173R, Zier2001A&A...377...23Z, Merritt2002Sci...297.1310M, Dennett-Thorpe2002MNRAS.330..609D}.
Whilst the spin-flip model encounters difficulty in explaining the subsequently discovered {\it inner} Z-symmetry of the wings, the model can still be tenable if the two jets are (temporarily) bent in opposite directions via interaction with the ISM of the parent galaxy, which is already set in rotation due to infall of a minor galaxy \citep{Gopal-Krishna2003ApJ...594L.103G}. 
A third scenario invokes a temporary diversion of the jet plasma due to jet-shell interaction \citep{Gopal-Krishna1983Natur.303..217G, Gopal-Krishna2012RAA....12..127G,Joshi2019ApJ...887..266J}.
Likewise, on much larger scale of 100s of kiloparsec, an XRG-shaped distortion of the lobes could arise from interaction of their backflow with the warm gaseous filaments of the cosmic web \citep{Gopal-Krishna2000ApJ...529..189G, Gopal-Krishna2009NewA...14...51G}, evidence for which has been reported recently \citep{Dabhade2022A&A...660L..10D}.
Clearly, in all these models the wings are expected to be filled with older synchrotron plasma than the primary lobes and hence they are generally not expected to exhibit a radio spectrum statistically indistinguishable from that of the primary lobes. 

A radically different model proposed for XRGs posits that the pairs of primary lobes and wings are being fuelled simultaneously by two separate bi-polar jets emanating from an unresolved binary of active SMBHs located within the nucleus of the parent galaxy \citep{Lal2005MNRAS.356..232L, Lal2019AJ....157..195L}.
Thus, in this `dual AGN' model, both lobe pairs are concurrently `active', leaving no reason to expect a systematic difference between their radio spectra.
This verifiable prediction contrasts with the prediction of the other XRG models mentioned above, according to which 
spectrum of a wing should be at least as steep as that of the primary lobe associated with that wing and fuelling it. Rare counter-examples to this straight-forward expectation, namely the XRGs 3C 223.1 and 3C 403, were first noticed by \citet{Rottmann2001PhDT.......173R}, these XRGs being the youngest members of his sample of 9 XRGs. The anomalous spectral distribution was confirmed for 3C 223.1, albeit only marginally, by \citet{Dennett-Thorpe2002MNRAS.330..609D} and also 
by \cite{Lal2005MNRAS.356..232L}.
Recently, a high-precision spectral index map of 3C 223.1, based on LOFAR (144 MHz) and VLA maps (4.9 GHz and 8.3 GHz) has been published, which demonstrates that, compared to the primary lobes, the spectrum is flatter in the wings and in certain parts of the wings it becomes even flatter than the spectrum of the terminal hot spots of the primary lobes \citep{Gopal-Krishna-Dhabade_2022A&A...663L...8G}. These authors have interpreted such rare, anomalous spectral pattern as an extreme case of in-situ particle acceleration probably associated with the bending of the streams of magnetised relativistic plasma backflowing from the hot spots into the associated wings.
Note that in recent relativistic  magneto-hydrodynamic simulations of XRGs, patches of flatter radio spectrum are seen to arise due to diffusive shocks inside the wings \citep{Giri2022A&A...662A...5G}. Nonetheless, from observational standpoint, a key question posed by these observations is whether this (rather counter-intuitive) spectral pattern is just a rare occurrence, or a fairly common feature of XRGs, as asserted in \citet{Lal2019AJ....157..195L}, based on their sample of 28 XRGs for which they had mapped the distribution of metre-wavelength spectral index (240 - 610 MHz). Given that spectral steepening is expected to be more pronounced at higher frequencies, a more  convincing approach would be to use radio imaging data that not only span a large frequency range but also extend as much beyond 1 GHz as possible.
The present study is a step in that direction.   
Here, we have combined the 1.4 GHz radio maps from the FIRST survey \citep{Becker_first_1995ApJ...450..559B} with the 144 MHz images from the LOFAR Two-meter Sky Survey Data Release 2 (LoTSS-DR2) \citep{Shimwell2022lotssdr2}, to accurately determine the spectral index distribution for a large unbiased sample of 24 XRGs, with the principal objective of verifying the above-mentioned dual-SMBH hypothesis for the origin of XRGs. To augment this sample, we have included one XRG, J1054+4703, taken from the uGMRT database (section \ref{sec:sample}). The selection of our XRG sample, data reduction and analysis details are provided in sections \ref{sec:sample} and \ref{sec:anlys}. In section \ref{sec:res}, our results are summarized and discussed briefly. The main conclusions are presented in section \ref{sec:con}. The spectral index maps for all the 25 XRGs are contained in Appendix \ref{sec:appndx} (see, also, Figure 1).

\section{The sample selection}
\label{sec:sample}
\begin{table} 
\begin{centering}
\scriptsize
\caption{\bf Sample selection}
\label{sample-selection}
\begin{tabular}{l l l }
\hline
XRG catalogue  & No. of XRGs    & Selection criteria adopted               \\
\hline
\citet{Cheung2007AJ....133.2097C}         & 100             &  $\Theta_{major}$$\ge$ 15\arcsec, dyn. range $\ge$ 1:40          \\
\citet{Proctor2011ApJS..194...31P}        & 156 (135 new)   &   see sect. \ref{sec:sample}                                                      \\   
\citet{Yang2019ApJS..245...17Y}	          & 290 (265 new)   &  $\Theta_{minor}$$\ge$ 10\arcsec, dyn. range $\ge$ 1:33          \\ 
\citet{Bera2020ApJS..251....9B}           & 161 (140 new)   &   $\Theta_{major}$$\ge$ 10\arcsec                                   \\   
 \hline	   
\end{tabular}
\end{centering}
\end{table}

Our main sample of 24 XRGs has been derived from published works \citep{Cheung2007AJ....133.2097C,Proctor2011ApJS..194...31P,Yang2019ApJS..245...17Y,Bera2020ApJS..251....9B}, which themselves are based on the 1.4 GHz VLA FIRST survey\footnote {\href{http://sundog.stsci.edu}{http://sundog.stsci.edu}} \citep{Becker_first_1995ApJ...450..559B}. Firstly, \citet{Cheung2007AJ....133.2097C} compiled a sample of 100 XRG candidates out of $1648$ targets, by applying the criterion of the radio major axis being $\ge$ 15\arcsec. 
\citet{Proctor2011ApJS..194...31P}, by implementing an automated morphological classification scheme over $811,117$ 
FIRST radio sources, identified 156 XRG candidates. Out of these, 21 XRGs  were already included in the \citet{Cheung2007AJ....133.2097C} sample, resulting in 135 new XRG candidates. Further, \citet{Yang2019ApJS..245...17Y} investigated 5128 FIRST radio sources having minor axis $ \ge 10$\arcsec and a dynamic range of $\ge 33:1$, and thus found 290 new XRGs, out of which 25 are already in the list of \citet{Proctor2011ApJS..194...31P}. They classified 106 of these 290 XRG candidates as strong XRG candidates and 184 as probable XRG candidates.
Lastly, \citet{Bera2020ApJS..251....9B} have recently catalogued 296 winged radio sources, drawn from the FIRST database containing 95,243 radio sources, by imposing a lower limit of $10\arcsec$ on the largest radio size. They classified 161 of these 296 winged radio sources as XRG candidates,
of which 21 are already contained in the \citet{Proctor2011ApJS..194...31P} catalogue. Thus, these above-mentioned samples of XRG candidates together resulted in our primary sample of 640 XRGs (see Table \ref{sample-selection}).

\begin{table*}   
\centering
\caption{\bf The XRG Sample}
\label{source-list}
\begin{tabular}{l c c c l r c l}
\hline
Name       &    RA       &    Dec      &    $z$    &  $z$-type  &    NVSS Flux      &  FIRST Flux &    Reference       \\
           &   (J2000)   &  (J2000)    &         &           &    at 1.4 GHz (mJy)         & at 1.4 GHz (mJy)     &                \\
\hline           
J0750+2825 & 07 50 01.81 & +28 25 09.9 &   0.531 &    Photo &    765.9 $\pm$ 26.4 &    750.5   &     \citet{Yang2019ApJS..245...17Y}\\
J0805+4854 & 08 05 44.0  & +48 54 58.1 &         &          &     34.2 $\pm$ 1.4  &    ~34.0   &      \citet{Cheung2007AJ....133.2097C} \\
J0823+5812 & 08 23 33.47 & +58 12 11.0 &   0.778 &    Spec  &     67.1 $\pm$   2.4 &     ~97.5   &     \citet{Yang2019ApJS..245...17Y} \\
J0832+4301 & 08 32 42.55 & +43 01 51.1 &   0.357 &    Photo &     59.5 $\pm$   2.2 &     ~58.7   &     \citet{Yang2019ApJS..245...17Y} \\
J0846+3956 & 08 46 03.58 & +39 56 57.9 &         &          &    196.7 $\pm$   6.7 &    197.1   &     \citet{Cheung2007AJ....133.2097C} \\
J0915+3401 & 09 15 24.03 & +34 01 40.6 &         &          &    186.2 $\pm$   6.1 &    182.1   &     \citet{Proctor2011ApJS..194...31P}\\
J0952+4059 & 09 52 25.77 & +40 59 35.5 &   0.469 &    Spec  &     74.0 $\pm$   2.7 &     ~77.6   &     \citet{Yang2019ApJS..245...17Y} \\
J1015+5944 & 10 15 41.14 & +59 44 45.2 &   0.527 &          &    221.4 $\pm$  7.7 &    219.4   &     \citet{Cheung2007AJ....133.2097C} \\ 
J1022+5213 & 10 22 12.66 & +52 13 42.4 &         &          &    148.3 $\pm$  5.1 &    141.7   &     \citet{Bera2020ApJS..251....9B} \\
J1054+4703*& 10 54 26.39 & +47 03 27.4 &   0.430 &     Spec &     101.8$\pm$ 3.7 &  ~135.6*    &     \citet{Yang2019ApJS..245...17Y} \\
J1202+4915 & 12 02 35.10 & +49 15 31.7 &         &          &    103.8 $\pm$ 3.6 &    103.5   &     \citet{Cheung2007AJ....133.2097C} \\
J1206+3812 & 12 06 17.35 & +38 12 34.9 &   0.838 &          &    246.5 $\pm$  8.1 &    241.1   &     \citet{Cheung2007AJ....133.2097C} \\
J1211+4539 & 12 11 02.50 & +45 39 14.4 &         &          &    231.9 $\pm$  7.6 &   226.1   &      \citet{Cheung2007AJ....133.2097C} \\
J1211+5717 & 12 11 22.95 & +57 17 52.8 &   0.730 &    Photo &    215.1 $\pm$  7.6 &    217.3   &     \citet{Yang2019ApJS..245...17Y}   \\ 
J1243+6404 & 12 43 16.02 & +64 04 25.3 &         &          &    241.5 $\pm$  7.9 &    234.1   &     \citet{Yang2019ApJS..245...17Y}\\
J1306+4257 & 13 06 52.38 & +42 57 02.0 &   0.657 &    Spec  &    145.3 $\pm$  5.0 &    143.1   &     \citet{Yang2019ApJS..245...17Y} \\
J1323+4153 & 13 23 09.94 & +41 53 18.0 &         &          &     76.9 $\pm$  2.7 &     ~74.6   &    \citet{Proctor2011ApJS..194...31P}\\
J1336+4313 & 13 36 36.06 & +43 13 29.0 &         &          &    272.4 $\pm$  9.4 &    272.7   &     \citet{Yang2019ApJS..245...17Y}\\
J1434+5906 & 14 34 02.17 & +59 06 53.3 &         &          &    315.0 $\pm$ 10.8 &    308.7   &     \citet{Cheung2007AJ....133.2097C}\\
J1537+3902 & 15 37 49.51 & +39 02 37.6 &         &          &     89.0 $\pm$  3.2 &     ~89.7   &     \citet{Bera2020ApJS..251....9B}\\
J1548+4451 & 15 48 17.19 & +44 51 47.4 &  0.599  &    Photo &    116.7 $\pm$  4.2 &    118.6   &     \citet{Yang2019ApJS..245...17Y}\\
J1558+3404 & 15 58 31.84 & +34 04 44.0 &  0.51   &          &     87.1 $\pm$  3.1 &     ~83.2   &     \citet{Bera2020ApJS..251....9B}\\    
J1601+3132 & 16 01 58.11 & +31 32 25.1 &  0.545  &    Photo &    136.3 $\pm$  4.8 &    134.9   &     \citet{Yang2019ApJS..245...17Y}\\
J1606+4517 & 16 06 38.88 & +45 17 37.1 &         &          &    115.6 $\pm$  4.0 &    115.9   &     \citet{Cheung2007AJ....133.2097C}\\  
J1633+3025 & 16 33 23.72 & +30 25 00.4 &  0.57   &          &    124.3 $\pm$  4.4 &    119.3   &     \citet{Bera2020ApJS..251....9B}\\
\hline
\end{tabular}
\\ 
* We have used the 1.4 GHz uGMRT image instead of the 1.4 GHz FIRST image (section \ref{sec:sample}) and also listed uGMRT fluxdensity in column 7 instead of the FIRST flux.
\end{table*}


We next cross-matched the above $640$  XRGs with the LoTSS-DR2\footnote  {\href{https://lofar-surveys.org/dr2_release.html}{https://lofar-surveys.org/dr2\_release.html}} \citep{Shimwell2022lotssdr2} catalogue containing 4,396,228 radio sources. The LoTSS-DR2 is a sensitive, high-resolution imaging survey made at 120--168 MHz, covering 27\% of the northern sky in two regions centered approximately at  
(RA = 12h45m, Dec = +44$^{\circ}$30\arcmin) \& (RA = 01h00m, Dec = +28$^{\circ}$00\arcmin), and spanning 4178 and 1457 square degrees, respectively. The survey has a 
median sensitivity of 83 $\mu$Jy beam$^{-1}$ (rms) and a beamsize (FWHM) of $\sim$ 6$^{\prime\prime}$. 
We found that 175 of the total 640 XRG candidates (originally selected from the 1.4 GHz FIRST catalogue, see above) are listed in the 144 MHz LoTSS-DR2 catalogue.  The similar angular resolutions of the 1.4 GHz FIRST ($\sim 5^{\prime\prime}$ ) and the 144 MHz LoTSS-DR2 ($6^{\prime\prime}$) surveys make them well-suited for spectral index mapping of our XRG sample. 
The LoTSS observations have a smaller uv lower limit (100 metre) than the FIRST observations. Consequently, the 144 MHz images have a higher sensitivity to any existing large-scale emission. We have attempted to minimise possible impact of this difference by ensuring compatibility of the FIRST (1.4 GHz) flux density with that measured in the NVSS (1.4 GHz) survey, which has a much greater sensitivity to extended structure. Through this selection filter, we have ensured that any `missing flux’ in the FIRST maps of the selected XRGs is insignificant. Note that we are interested here primarily in spectral gradients and not in the absolute values of spectral index for different regions of our sources.
Thus, we have only selected those sources whose quoted 1.4 GHz FIRST flux density falls within 1$\sigma$ error of (or 
exceeds) that reported in the 1.4 GHz NRAO VLA Sky Survey (NVSS) which is made with a much larger beam of $45^{\prime\prime}$ diametre and is known to be sensitive to diffuse radio emission of extent up to at least 10 arcmin \citep{Condon_nvss_1998AJ....115.1693C}. 
This cautionary approach reduced our sample from 175 to 41 XRG candidates deemed suitable for spectral imaging. 
Finally, we scrutinized the morphologies of all these 41 sources in the 3.0 GHz VLA Sky Survey\footnote {\href{https://science.nrao.edu/vlass}{https://science.nrao.edu/vlass}} (VLASS \citep{Lacy-vlass-2020PASP..132c5001L}), 1.4 GHz FIRST and the 144 MHz 
LoTSS-DR2 surveys, in order to confirm that in each source at least one wing is clearly delineated. This procedure left us with a sample of 24 XRGs. 
To augment this sample, we have included one XRGs, J1054+4703, obtaining its map from our data taken with the upgraded Giant Metrewave Radio Telescope (uGMRT, \citep{Swarup1991CSci...60...95S, Gupta2017CSci..113..707G}) and then generated its spectral index map by combining the 1.4 GHz uGMRT map with the 
144 MHz LoTSS-DR2 map. 
We have preferred the available uGMRT map over its FIRST map at the same frequency of 1.4 GHz, in view of its superior uv coverage, particularly at short spacings, as well as its higher sensitivity. Thus, with the addition of this source, we arrived at a final sample of 25 XRGs (Table \ref{source-list}).
The rms errors in the FIRST images of our sample vary between 125 to 175 $\mu$Jy beam$^{-1}$, while for the LoTSS DR2 images, the corresponding range is 160 to 483 $\mu$Jy beam$^{-1}$. For our uGMRT image, the rms error is 129 $\mu$Jy beam$^{-1}$. We emphasize that in the entire sample selection process, we have taken no cognisance of any available radio spectral information for the sources.

\section{Data Analysis}
\label{sec:anlys}
For analysis of the 1.4 GHz FIRST/uGMRT maps and the 144 MHz LoTSS-DR2 maps of our sample of 25 XRGs, we used the Astronomical Image Processing System 
(AIPS)\footnote{\href{http://www.aips.nrao.edu/}{http://www.aips.nrao.edu/}}. Firstly, 
the 1.4 GHz maps of the XRGs were smoothed to the $6\arcsec$ FWHM of their LoTSS DR2 maps, using the AIPS task CONVL. 
The geometry of the images at the two frequencies was then adjusted, in order to match them in pixel sampling, coordinates and reference pixels, using the AIPS task HGEOM. 
Adopting the convention  $S \propto \nu^{\alpha}$, 
we then produced the spectral index ($\alpha$) map for each XRG by calculating $\alpha$ for each pixel in the maps, using the AIPS task COMB. 
For this purpose, we have only considered the regions of the VLA FIRST and LoTSS-DR2 
maps within their 5$\sigma$ contours in order to keep the uncertainties on $\alpha$ small. 
Comparison of the LOFAR map with the FIRST map smoothed to the 6 arcsec circular beam of the LOFAR map, yielded pixel-by-pixel distribution of the spectral index across a given XRG. The rms error on $\alpha$ was computed using the relation:
\vskip 0.1cm
$\delta\alpha= \frac{1} {ln(\nu_1/\nu_2)} \sqrt{(\frac{\sigma S_1}{S_1})^2+(\frac{\sigma S_2}{S_2})^2}$ \\
\vskip 0.1cm
where $\sigma S$ is the rms error on flux density, $S_1$ and $S_2$ are the flux densities at frequency $\nu_1$ and $\nu_2$, respectively.

To determine the spatial variation of spectral index for our XRG sample, we drew a profile path (ridge-line) joining the brightest parts of the primary lobe and its associated wing.  In Fig. \ref{Fig-A}, we show the variation of spectral index, measured at between 3 to 8 circular regions along the extent of the profile path (marked as a contiguous sequence of circles). Thus, $\alpha$ computed at each point along the profile path corresponds to the brightness averaged over a circular region of $6^{\prime\prime}$ diameter, being the image resolution. This approach is fairly conventional for measuring spectral index gradients in both classical double radio sources \citep[e.g., 3C 236,][]{Shulevski2019A&A...628A..69S} and XRGs \citep[e.g.,][]{Rottmann2001PhDT.......173R, Klein1995A&A...303..427K}.

\begin{figure*}
\begin{center}
\includegraphics[width=16 cm]{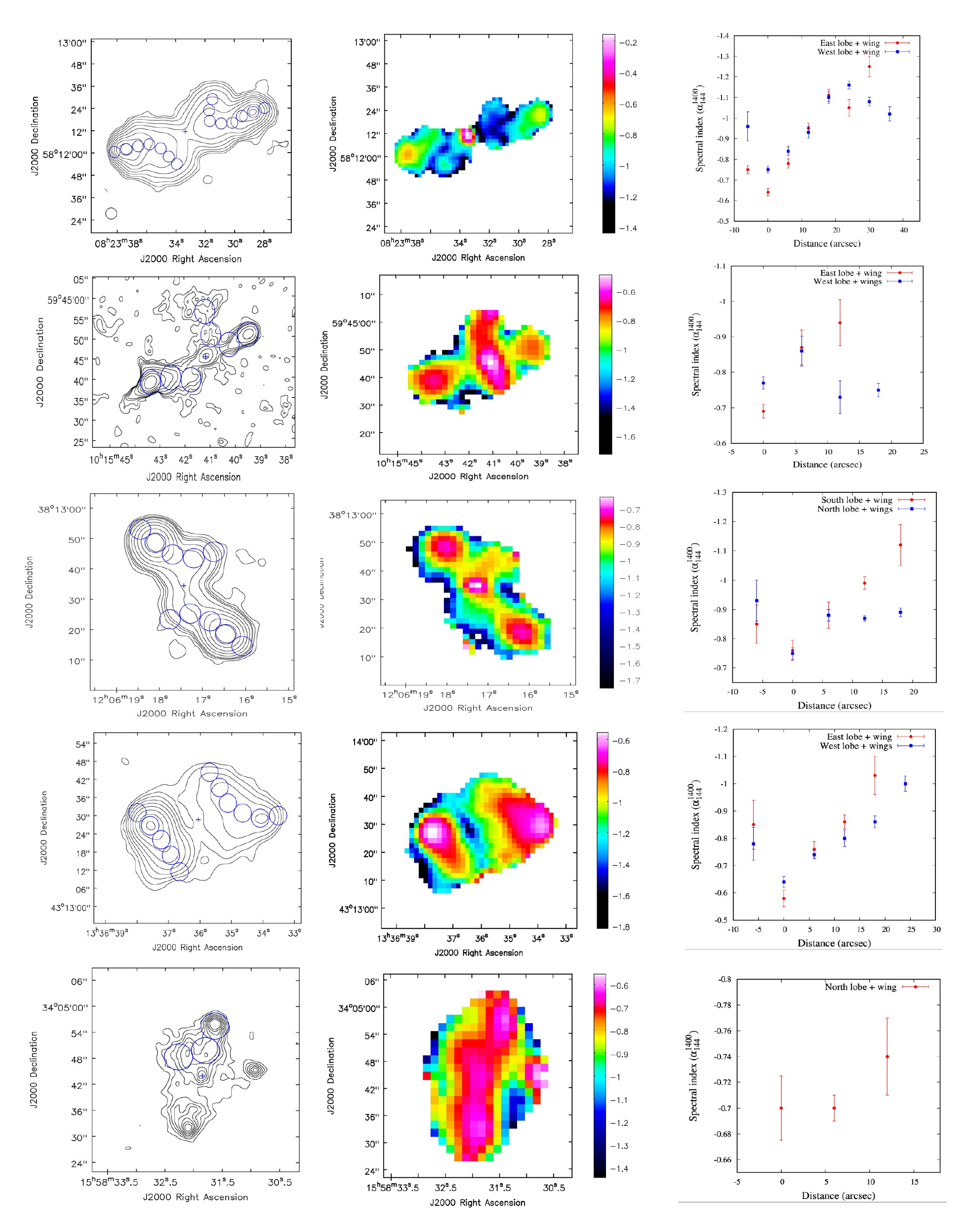}
\caption{Radio contour map, spectral-index image (144 - 1400 MHz) and the corresponding spectral index profile measured along the rigde-line(s) marked as a contiguous chain of circles in the contour map, are displayed row-wise for 5 of the 25 XRGs, namely J0823+58, J1015+59, J1206+38, J1336+43 and J1558+34 (see text). The right panel shows the spectral index measured at the points marked with 6 arcsec diameter circles placed contiguously along the corresponding ridge-line(s)(left panel). 
In the right column, the origin along the x-axis coincides with the hot spot. The XRG nucleus is marked with a `+' sign in each contour map.} 
\label{Fig1}
\end{center}
\end{figure*}

\section{Results and Discussion}
\label{sec:res}

The derived spectral index distributions for all the 25 XRGs 
are displayed in Appendix A (Fig. \ref{Fig-A}), together with additional information derived from them (see below). The data for each XRG are presented in a single-column mosaic consisting of 6 panels, showing from top to bottom:
(i) VLASS (3.0 GHz) map; (ii) FIRST (1.4 GHz) map\footnote{For J1054+4703 we have used its uGMRT map at 1.4 GHz, as discussed in sect. \ref{sec:anlys}.}; (iii) LoTSS-DR2 (144 MHz) map; (iv) map of the computed spectral index between 144 MHz and 1400 MHz ($\alpha^{1400}_{144}$), (v) spectral index error map, and (vi) the profile of $\alpha^{1400}_{144}$ running along the ridge-line(s) marked in one of the above 
three maps deemed best suited for identifying the wing component extending from the primary lobe. For illustration purposes, we show in Fig. \ref{Fig1} the mosaics for 5 of our XRGs, whose spectral properties are highlighted below.

Our principal objective here is to verify the remarkable conclusion reached in \citet{Lal2019AJ....157..195L}, based on their analysis of a sample of 28 XRGs, that there is essentially no systematic difference between the radio spectra of the primary lobes and their associated wings, all imaged by them at metre wavelengths (610 \& 240 MHz). Such a check is important because this conclusion runs counter to the expectation from nearly all the models proposed for XRGs, including the two most
prominent ones, namely the `spin-flip' and the `backflow diversion' models (sect. \ref{sec:intro}). We reckon the approach followed here to be superior to that followed in \citet{Lal2019AJ....157..195L} in two respects. Firstly, their spectral index estimates refer to a rather narrow frequency range (240 - 610 MHz) which, moreover falls within the sub-GHz range where spectral steepening due to radiative aging is likely to be small (even negligible). In contrast, the spectral index values determined here are between 144 MHz and 1400 MHz, thus covering an order-of-magnitude frequency ratio and hence conducive to greater precision. Secondly, the method of measuring spectral index continuously along a ridge-line is less subjective than measuring it at some chosen locations within the source structure, as done in \citet{Lal2019AJ....157..195L}.
For 40 out of the total 50 primary lobes in our sample of 25 XRGs, was it possible to mark the ridge-line with reasonable confidence and the spectral index profiles along those 40 ridge-lines running through the primary lobes and their respective wings, are displayed in Fig. \ref{Fig-A}  (although spectral index maps are presented for all 25 XRGs). The same data for a subset of 5 of the XRGs are displayed in Fig. \ref{Fig1}. 
We note that for 3 of the 40 lobe-wing structures, namely those in the XRGs J0750+2825, J1211+5717 and J1558+3404, the wing is only marginally resolved from the primary lobe. Nonetheless, their wings encompass at least two sampled points (shown as blue circles in Figure \ref{Fig-A}).

It may further be mentioned that the spectral index estimation along a ridge-line should be a more reliable indicator of the spectral trend also because a ridge-line traces the brightest parts of the lobe and its associated wing. Clearly, a spectral difference between the primary lobe and the wing (which signifies spectral ageing, however, see below) should be more readily discernible in such ridge-line $\alpha$ profiles (section \ref{sec:anlys}).
Recall that the same method was used in the original study of $\alpha$-distribution in XRGs, which led to the unexpected discovery of a distinctly flatter spectrum in the wings of an XRG (3C 223.1), compared to its primary lobes \citep{Rottmann2001PhDT.......173R}. As noted in section \ref{sec:intro}, this anomalous trend has recently been confirmed using much deeper radio maps of this XRG made with superior angular resolution at a pair of frequencies differing by a huge factor of 57, which imparts a high precision to spectral index determination \citep{Gopal-Krishna-Dhabade_2022A&A...663L...8G}. 
Thus, XRGs with such an abnormal spectral gradient do exist, underscoring the need for {\it in-situ} particle acceleration occurring in the wings. But, are such spectral gradients mere anomalies, or a fairly common feature of XRGs, as concluded in \citet{Lal2019AJ....157..195L}, is the key question being investigated here.  

Going through Figure \ref{Fig-A}, the  most common trend clearly is for the spectrum to steepen along the ridge-line from the hot spot (primary lobe) towards the 
associated wing. This is consistent with both leading models of XRGs, namely the `backflow diversion' and the `spin-flip' models (Sect. \ref{sec:intro}).
{\it For neither of the 40 ridge-lines do we clearly observe a spectral flattening towards the wing, the single possible exception being the northen lobe of XRG 
J1015$+$59}. Below we note the four interesting spectral trends discerned in our sample, examples of which are displayed in Fig. \ref{Fig1}:

\begin{enumerate}

\item Along several ridge-lines, the spectral index is found to remain constant (i.e., no spectral gradient). 
They belong to the XRGs J1015+5944 (W-lobe+wing), J1323+4153 (E-lobe+wing, possible case), J1558+3404 (N-lobe+wing), J1601+3132 (N-lobe+wing) and J1633+3025 (both lobes+ their wings).

\item In the XRG J0823+5812 (W-wing) $\alpha$ appears to flatten towards the wing's extremity.
If this `near-the-edge' spectral flattening is confirmed with more sensitive observations, that would strengthen the case for $\it in~situ$ particle acceleration taking place near the extremities of these wings.  Further, we may recall that the XRG J1015+5944 is one possible case in the present sample of 25 XRGs, whose (western)
lobe is suggestive of spectral flattening towards the associated (northern) wing. However, as seen from Fig. \ref{Fig1}, this needs confirmation through more sensitive and spatially better resolved radio imaging, possibly extending to a few GHz.

\item In several XRGs, significant radio emission is detected even beyond the hotspot (Fig. \ref{Fig-A}). Interestingly, some of these radio `spurs'  (or `plumes') extending from the hotpots appear to have a steeper spectrum than their respective hot spots. From Fig. \ref{Fig-A}, such primary lobes belong to the XRGs J0823+5812 
(both lobes), J0915+3401 (W-lobe), J1202+4915 (both lobes), J1206+3812 (both lobes) and J1336+4313 (both lobes). While this spectral index pattern shows that the hot spot is indeed the site of particle acceleration, it appears that at least in these cases, synchrotron plasma has not been fully diverted into the backflow from the hot spot, but has instead got accumulated in the region ahead of the hotspot. This raises the possibility that the advance of these hot spots may be (temporarily) stalled. 
A detailed investigation of this observed trend is clearly warranted. We recall that a similar spectral pattern has been noticed in the N-lobe of the giant radio galaxy 0506-286 (Figure 3 in \citep[][]{Dabhade2022A&A...660L..10D} and even in the classical double radio source Cygnus A \citep{Carilli1991ApJ...383..554C}. A more distant example is the $z = 3.4 $ radio galaxy B2 0902+34 \citep{Cordun2023arXiv230600071C}.

\item Adding to the spectral diversity of our XRG sample are the few XRGs where the lobes on opposite sides of the parent galaxy have distinctly different spectral gradients, albeit neither shows a clear spectral flattening towards the wings. Examples are J1015+5944 and J1206+3812 (Fig. \ref{Fig1}). 

\end{enumerate}

A major application of spectral gradients in radio galaxies has traditionally been for the purpose of estimating the (radiative) ages of their lobes.  This subject has a history going to 4-5 decades back, during which increasingly sophisticated theoretical formulations have been devised 
\citep[e.g.,][]{Burch1979MNRAS.186..519B,Winter1980MNRAS.192..931W,Myers1985ApJ...291...52M,Leahy1989MNRAS.239..401L,Carilli1991ApJ...383..554C,Katz-Stone1997ApJ...488..146K, Blundell2000AJ....119.1111B, Murgia2002NewAR..46..307M,Harwood2017MNRAS.469..639H}.
The well-known complicating factors in the age estimation include global processes occurring within the lobes, such as backflow mixing, adiabatic losses due to lobe expansion and {\it in-situ} particle acceleration. Additional possibilities include a changing injection spectrum of the radiating particles over the source lifetime
\citep[e.g.,][]{Leahy2002IAUS..199..179L}
and non-uniformity of the lobe's magnetic field. The latter could be on large scales 
\citep{Wiita1990ApJ...353..476W, Blundell2000AJ....119.1111B},
or on small-scales as well
\citep[e.g.,][]{Eilek1989AJ.....98..256E, Owen1989ApJ...340..698O, Siah1989BAAS...21.1094S,Tribble1993MNRAS.261...57T, Eilek1997ApJ...483..282E,Kaiser1997MNRAS.292..723K}.
A discussion on the import of such issues for age estimation of radio galaxies is presented, e.g., in \citet{Rudnick2002NewAR..46...95R} and \citet{Leahy2002IAUS..199..179L}.
In the specific case of XRGs, the role of such complicating factors has been highlighted in the relativistic magneto-hydrodynamic simulations, reported by \citet{Giri2022A&A...662A...5G}, in particular  the role of (diffusive) shock acceleration occurring within the wings and potentially even guiding their evolution. These authors also argue 
that the wings showing a flatter radio spectrum than the associated primary lobes would not necessarily imply a dual central engine 
(Sect. \ref{sec:intro}) and it could instead arise due to a suitable combination of viewing angle, evolutionary stage and the radio-frequency range adopted. The spectral imaging studies like the one reported here, when extended to bigger sample covering a larger frequency span, can provide a much needed observational feedback and constraints to the increasingly more sophisticated numerical simulation studies.

\section{Conclusions}
\label{sec:con}
We have derived spatial variation of spectral index for a well-defined unbiased sample of 25 XRGs, by combining their images at 1.4 GHz (VLA FIRST, or uGMRT) with their 144 MHz LoTSS-DR2 maps. For XRGs, this work represents the best combination, hitherto, of sensitivity, spatial resolution, frequency range, and sample size.
The most common trend found here is for the spectrum to steepen from the hot spot towards the wing along the ridge-line. This is consistent with both leading models of XRGs, namely the `backflow diversion’ and the `spin-flip’ models (sect. \ref{sec:intro}).
We find no evidence supporting the reported assertion for XRGs that the wings having a flatter spectrum than their associated primary lobes are 
just as common as the converse (sect. \ref{sec:intro}). Thus, although such a spectral gradient has actually been established for the XRG 3C 223.1 (sect. \ref{sec:intro}), it must be exceedingly rare and the only possible such case in the present sample is the XRG J1015+5944. For the XRG, J0823+5812 (W-lobe+wing), we find a
hint of spectral flattening towards the wing's extremity (sect. \ref{sec:res}). If confirmed by more sensitive observations, that would be an evidence for {\it in-situ} particle acceleration occurring in the outermost regions. Specifically for XRGs, such a possibility is indicated in the relativistic magneto-hydrodynamic simulations by \citet{Giri2022A&A...662A...5G}, from which they infer that re-energized particles due to shocks largely control the evolution of the wings \citep[see, also,][]{Chibueze2021Natur.593...47C}.

\section*{Acknowledgments}
\bibliographystyle{mnras}
We thank an anonymous reviewer for the helpful comments on the original version of the manuscript.
DP acknowledges the post-doctoral fellowship of the S.~N.~Bose National Centre for Basic Sciences, Kolkata, India, funded by the Department of Science and Technology (DST), India.
GK thanks Indian National Science Academy for a Senior Scientist position. GK and DP acknowledge the Indian Institute of Astrophysics for the hospitality during their visit when part of this work was carried out.

\section*{Data Availability}
We have used the VLASS, FIRST, and LoTSS DR2 data for our analysis in this manuscript. Appropriate links are given in the manuscript.
The $u$GMRT radio data used in this study will be publicly available at $u$GMRT data archive at \href{https://naps.ncra.tifr.res.in/goa/data/search}{https://naps.ncra.tifr.res.in/goa/data/search}.




\bsp	
\bibliography{references}
\appendix
\section{}
\label{sec:appndx}

\begin{figure*}
\begin{center}
\includegraphics[width=15 cm]{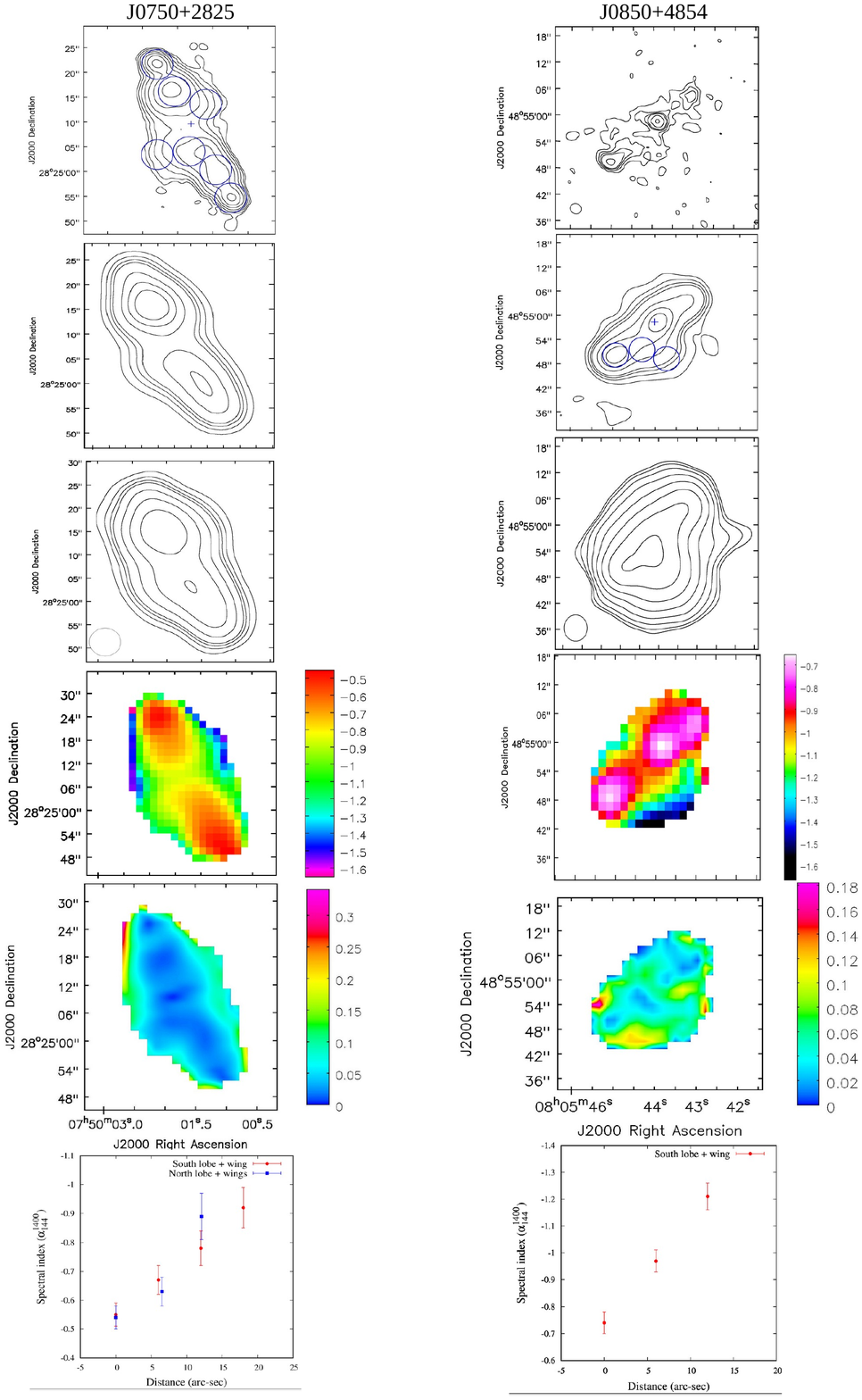}
\caption{The radio contour maps (VLASS, FIRST, and LoTSS DR2), spectral-index image (144 MHz -- 1.4 GHz), spectral-index error map, and the corresponding spectral index profile measured along the ridge-line (s) marked in the radio contour maps are displayed column-wise for our sample of 25 XRGs. The name of each source is mentioned at the top of each column. The bottom panel in each column shows values of spectral index at several points along the ridge-line(s). These values represent average taken over the circular regions of 6 arcsec diameter (the beamsize), as displayed in one of the first three maps deemed best suited for identifying the wing component. For each XRG, the location of the nucleus is marked on the contour map with a `+'.} 
\label{Fig-A}
\end{center}
\end{figure*}
\addtocounter{figure}{-1}
\begin{figure*}
\begin{center}
\includegraphics[width=16 cm]{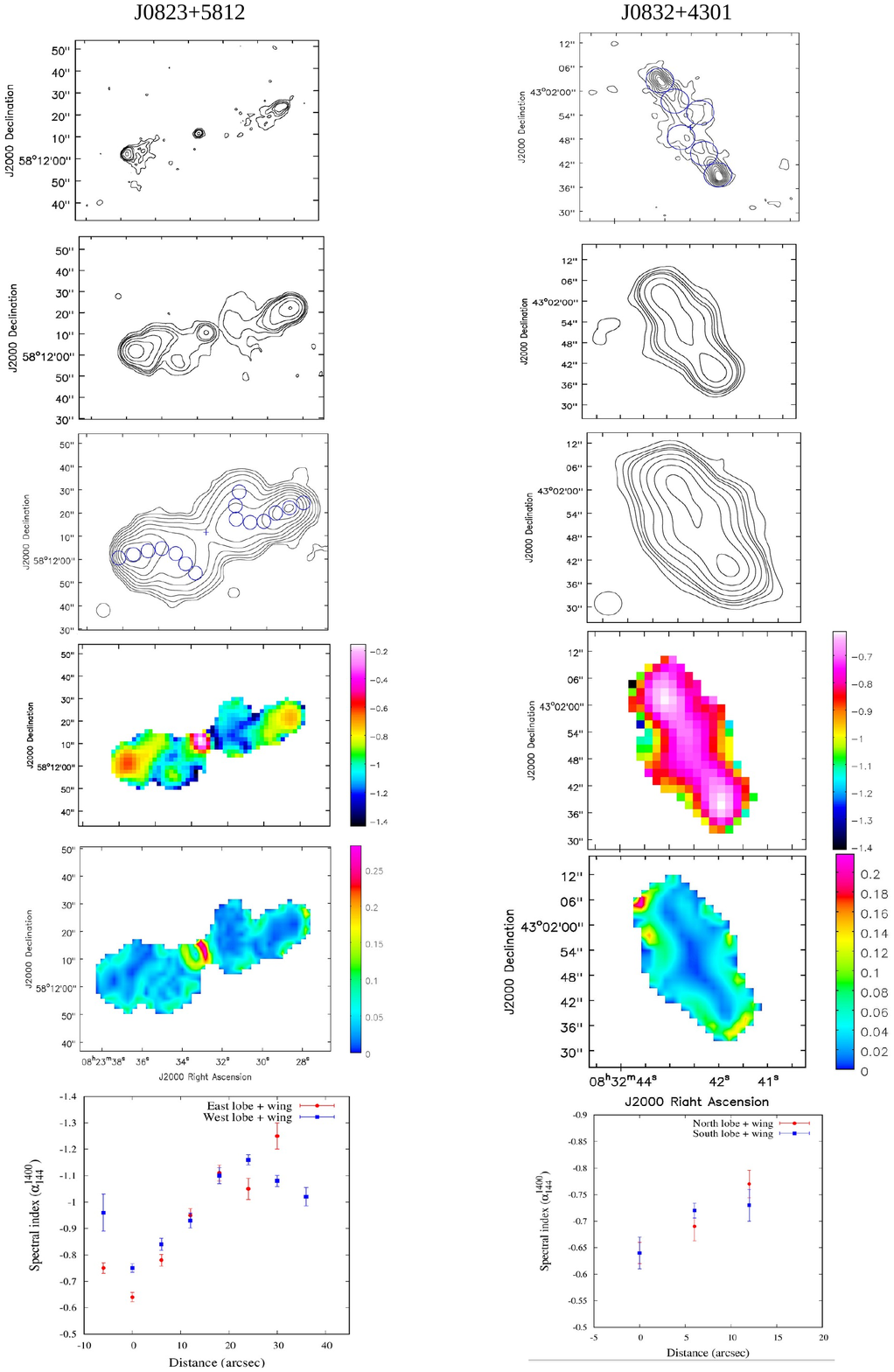}
\caption{Continued} 
\end{center}
\end{figure*}
\addtocounter{figure}{-1}
\begin{figure*}
\begin{center}
\includegraphics[width=16 cm]{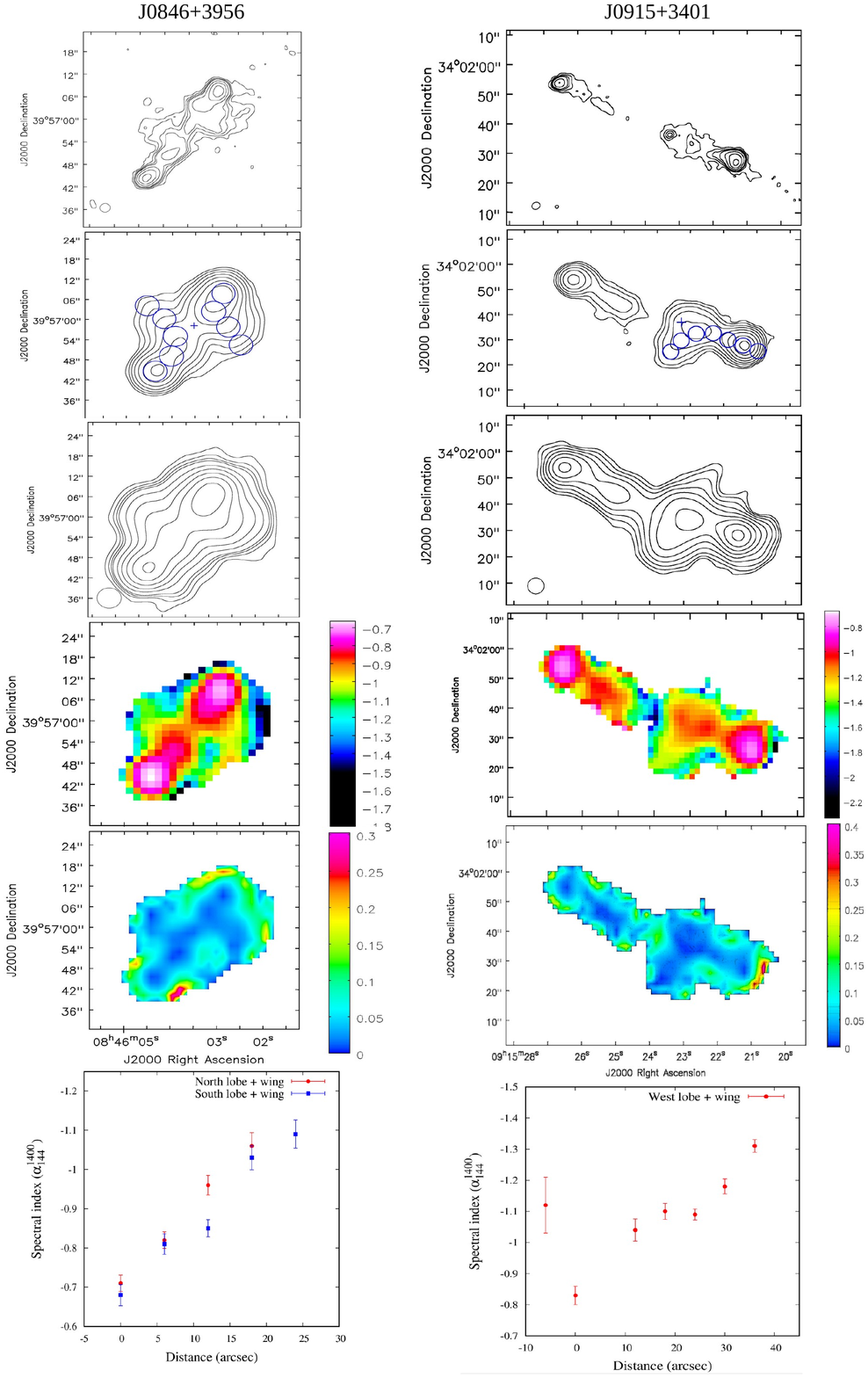}
\caption{Continued}  
\end{center}
\end{figure*}
\addtocounter{figure}{-1}
\begin{figure*}
\begin{center}
\includegraphics[width=16 cm]{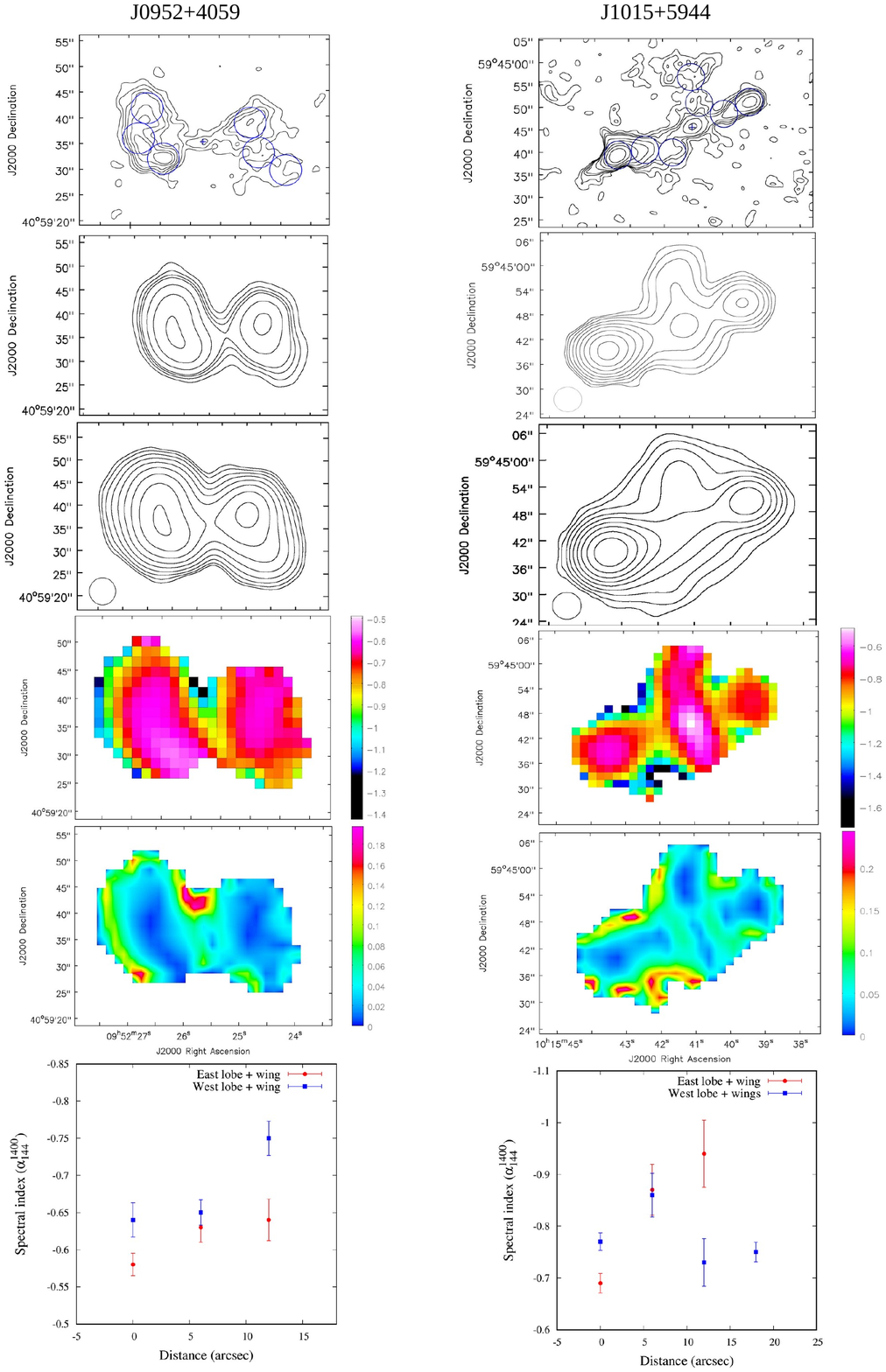}
\caption{Continued} 
\end{center}
\end{figure*}
\addtocounter{figure}{-1}
\begin{figure*}
\begin{center}
\includegraphics[width=16 cm]{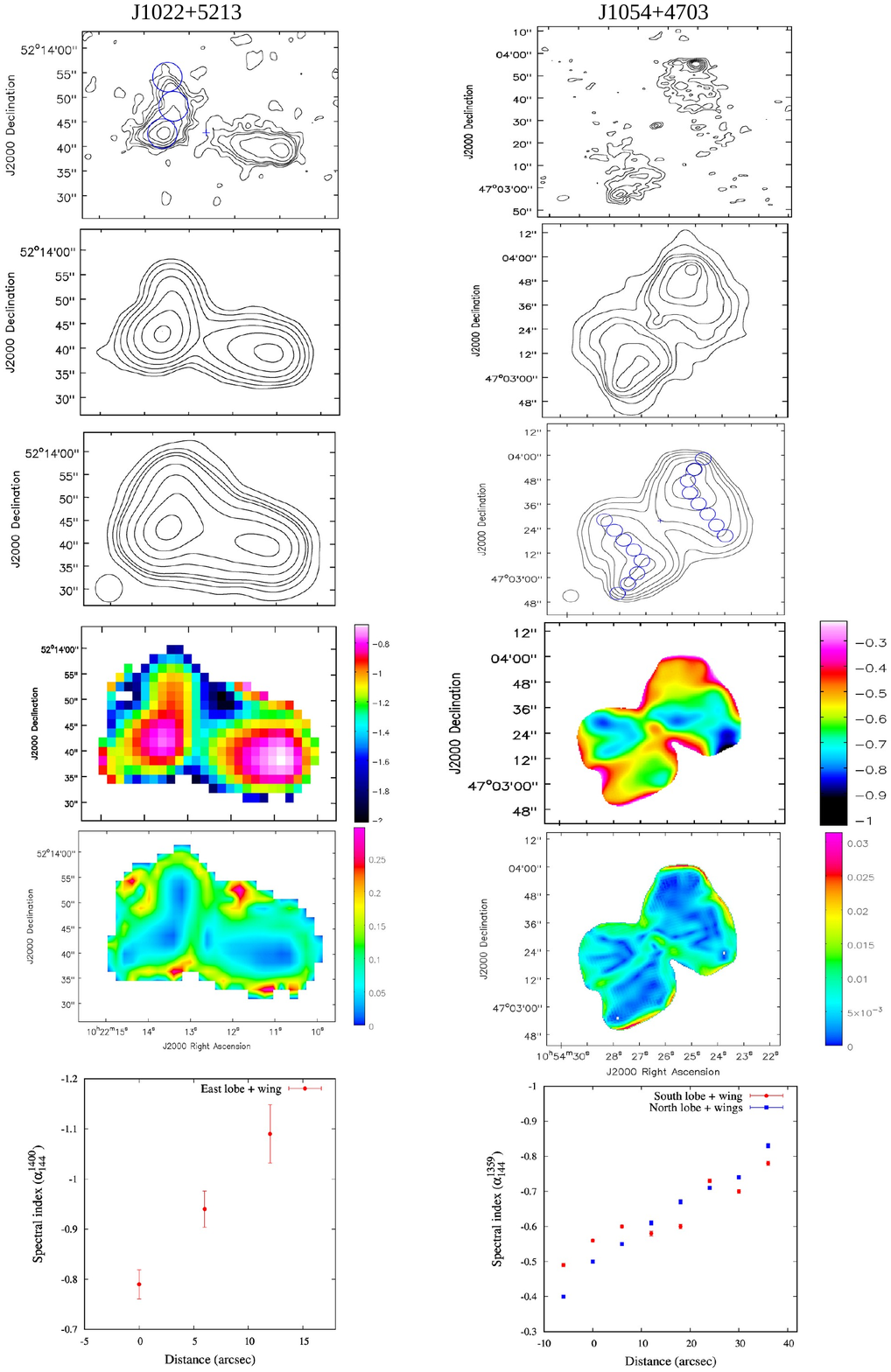}
\caption{Continued} 
\end{center}
\end{figure*}
\addtocounter{figure}{-1}
\begin{figure*}
\begin{center}
\includegraphics[width=16 cm]{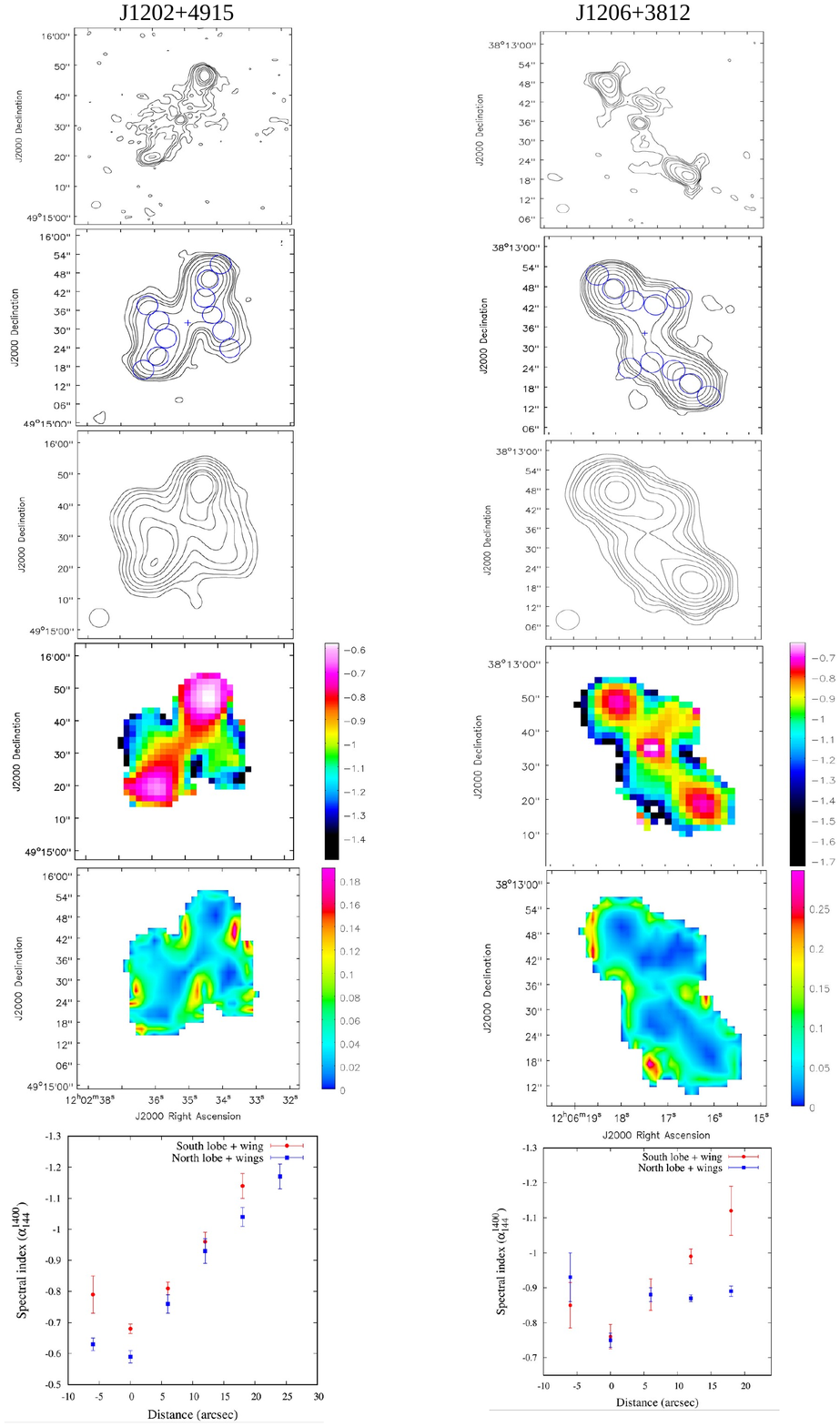}
\caption{Continued}  
\end{center}
\end{figure*}
\addtocounter{figure}{-1}
\begin{figure*}
\begin{center}
\includegraphics[width=16 cm]{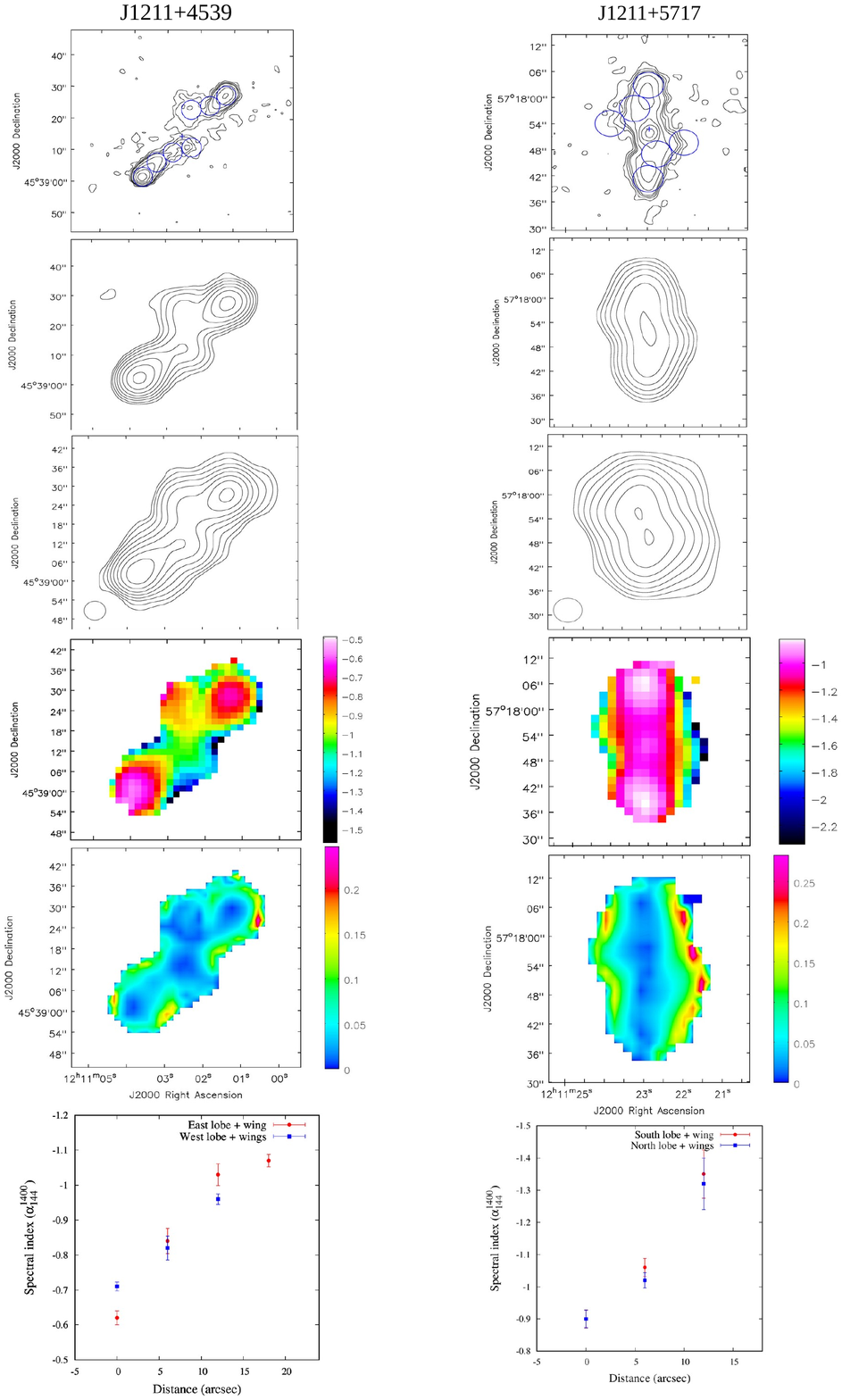}
\caption{Continued} 
\end{center}
\end{figure*}
\addtocounter{figure}{-1}
\begin{figure*}
\begin{center}
\includegraphics[width=16 cm]{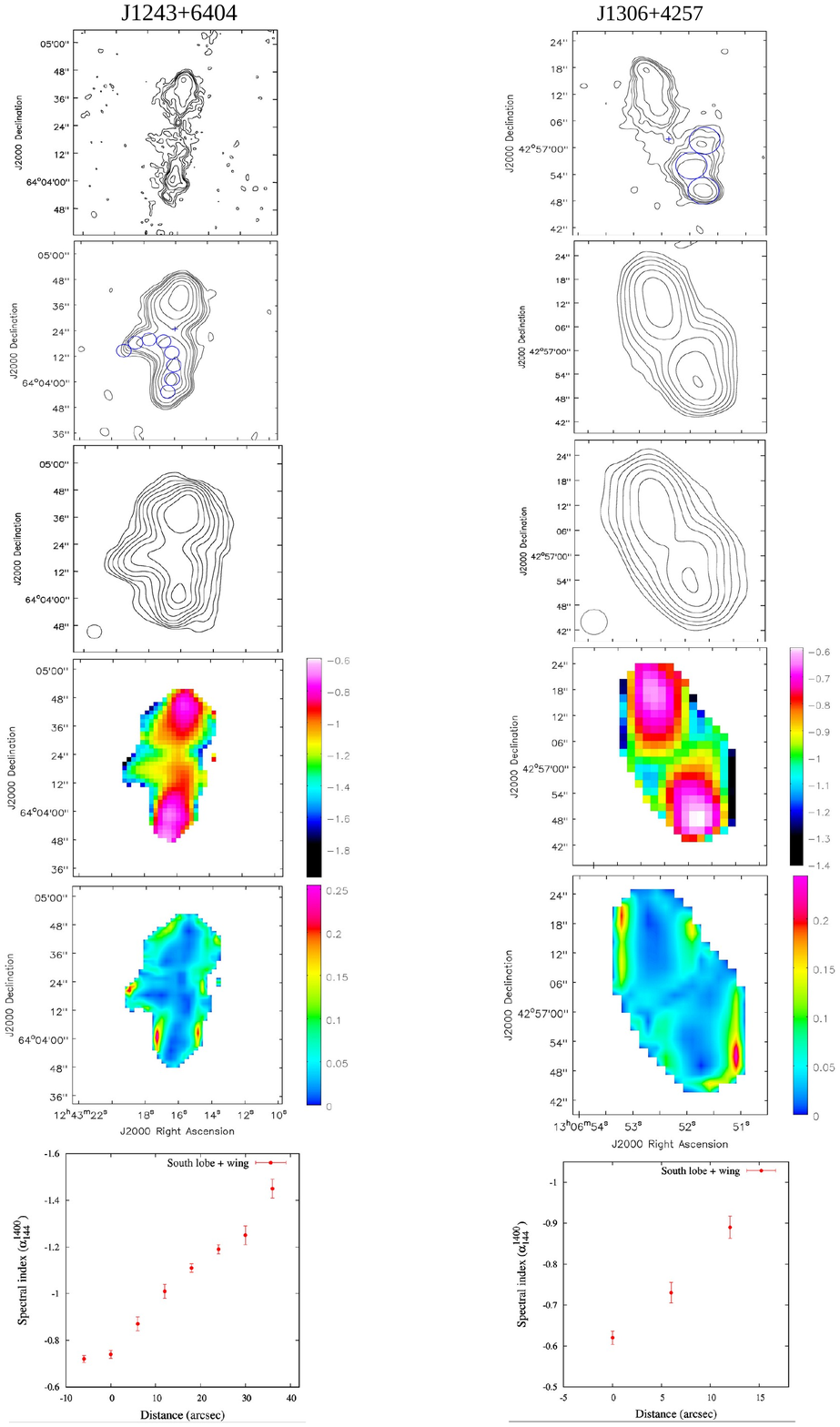}
\caption{Continued} 
\end{center}
\end{figure*}
\addtocounter{figure}{-1}
\begin{figure*}
\begin{center}
\includegraphics[width=16 cm]{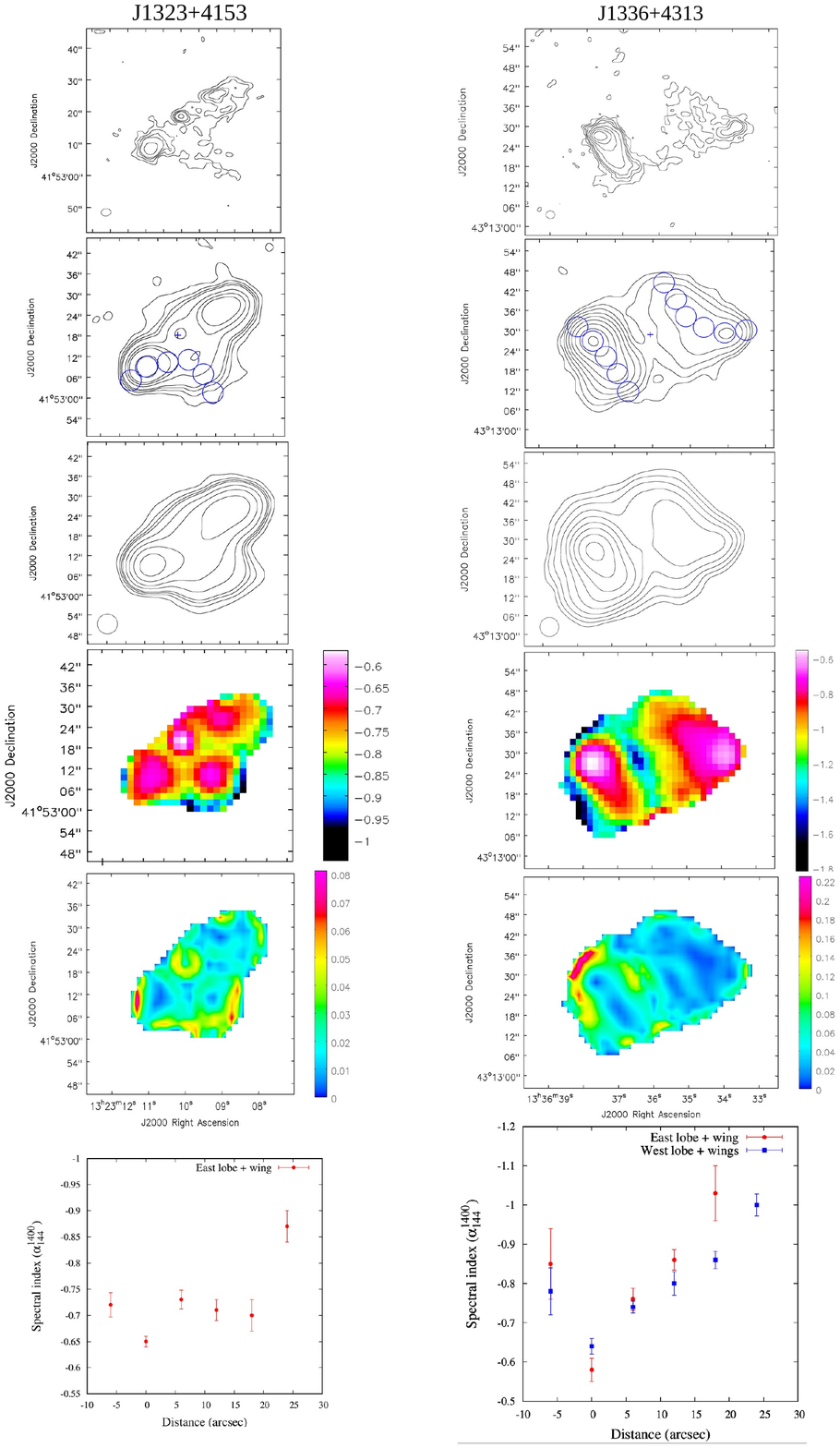}
\caption{Continued}  
\end{center}
\end{figure*}
\addtocounter{figure}{-1}
\begin{figure*}
\begin{center}
\includegraphics[width=16 cm]{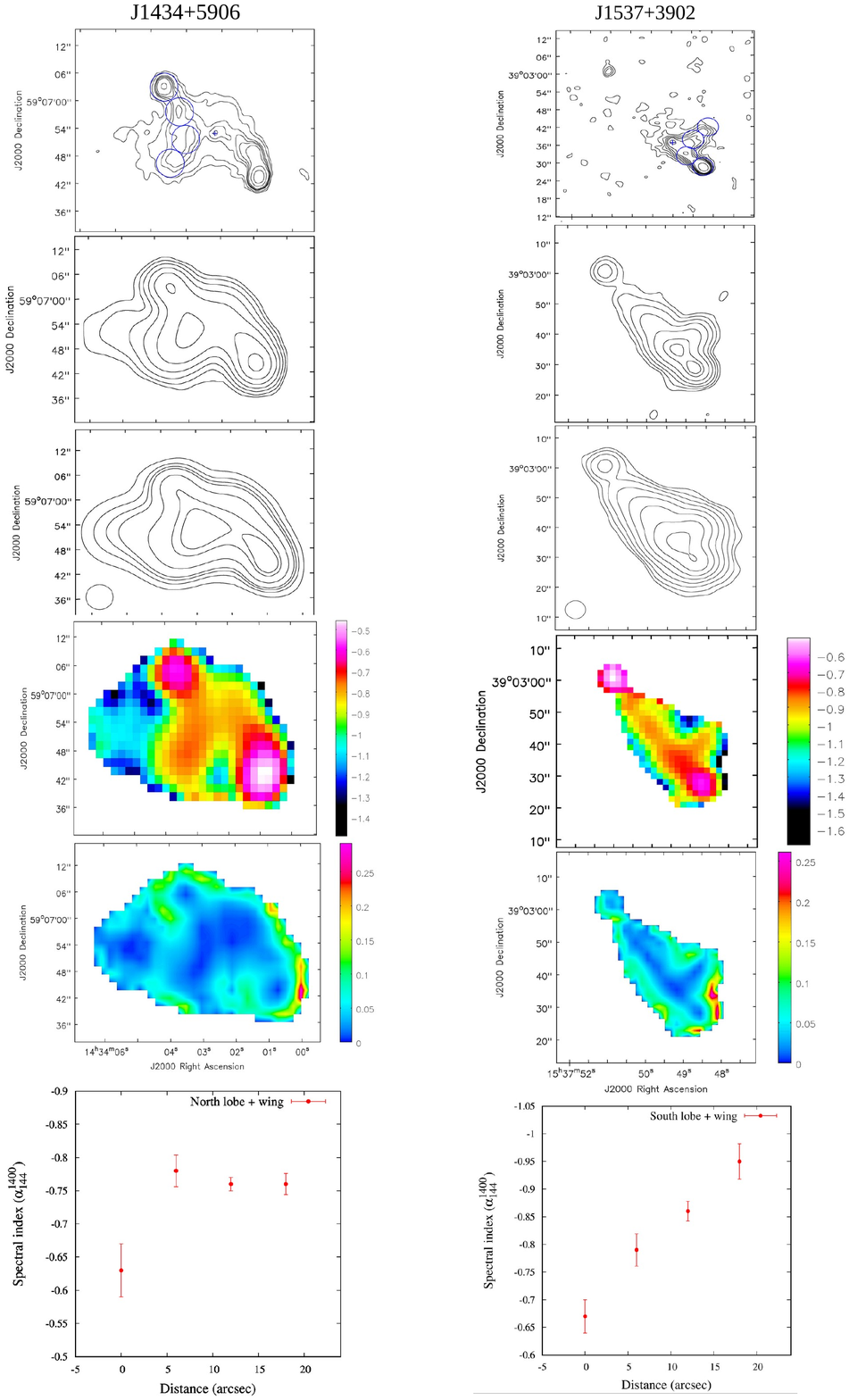}
\caption{Continued} 
\end{center}
\end{figure*}
\addtocounter{figure}{-1}
\begin{figure*}
\begin{center}
\includegraphics[width=16 cm]{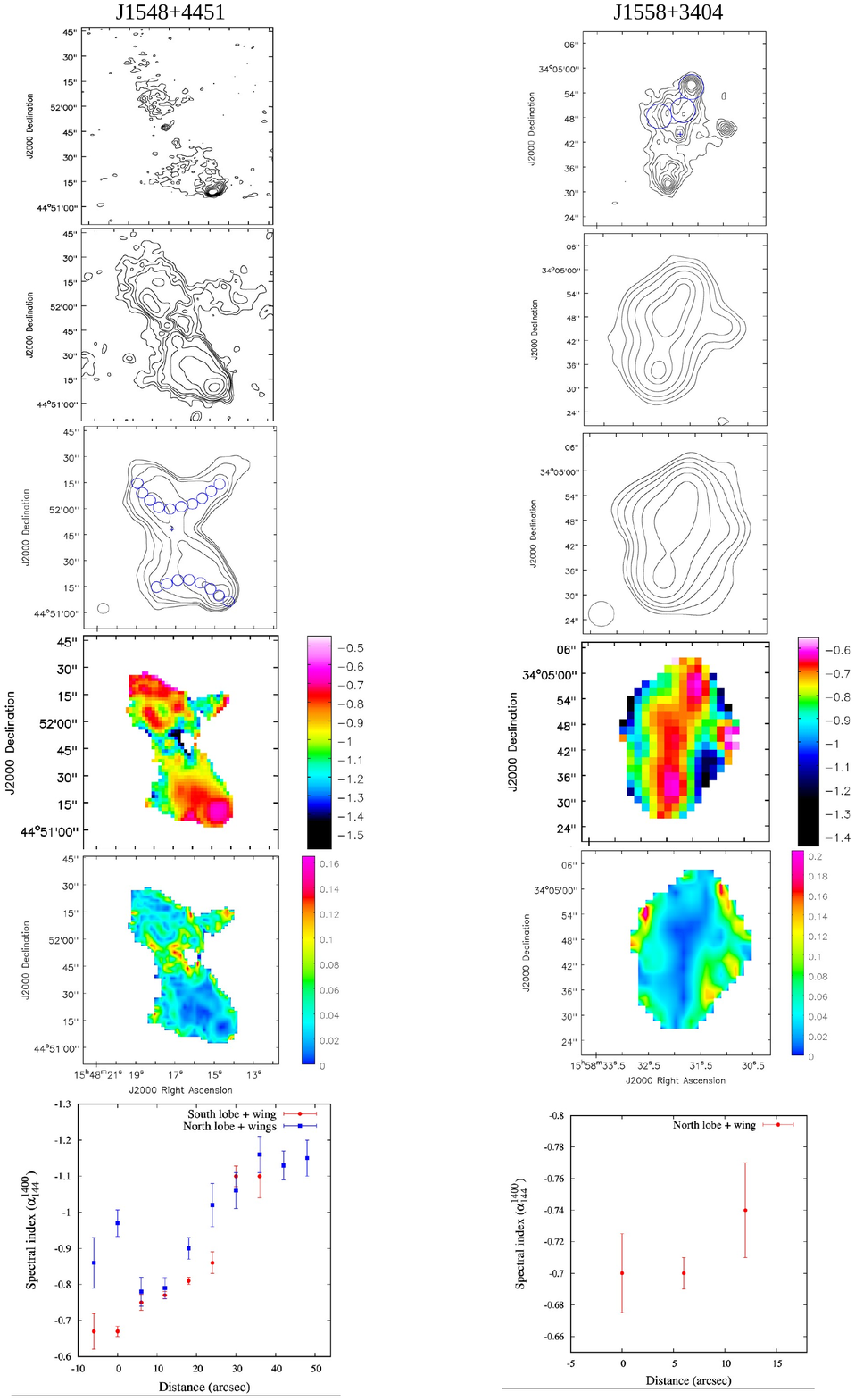}
\caption{Continued} 
\end{center}
\end{figure*}
\addtocounter{figure}{-1}
\begin{figure*}
\begin{center}
\includegraphics[width=16 cm]{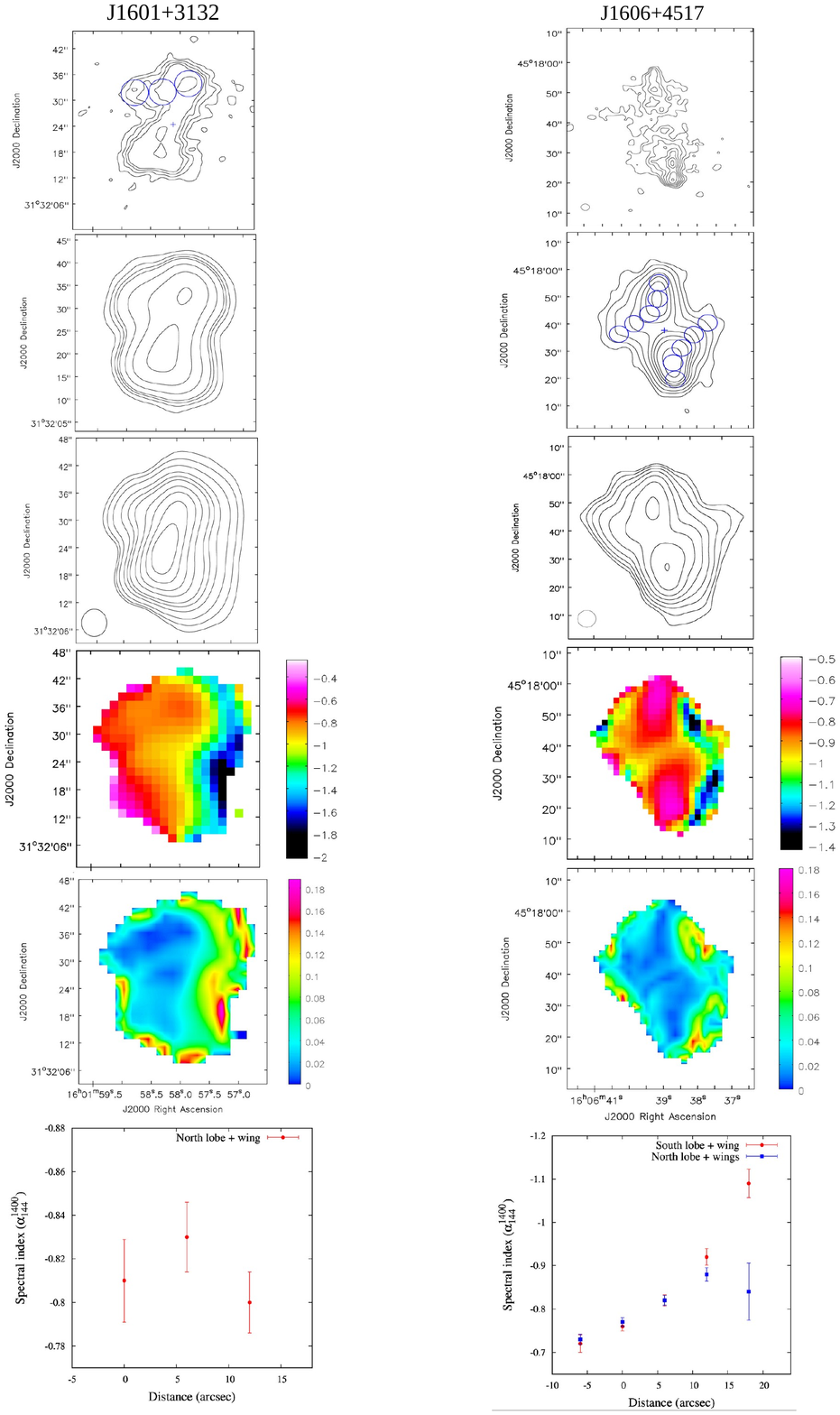}
\caption{Continued} 
\end{center}
\end{figure*}
\addtocounter{figure}{-1}
\begin{figure*}
\begin{center}
\includegraphics[width=16 cm]{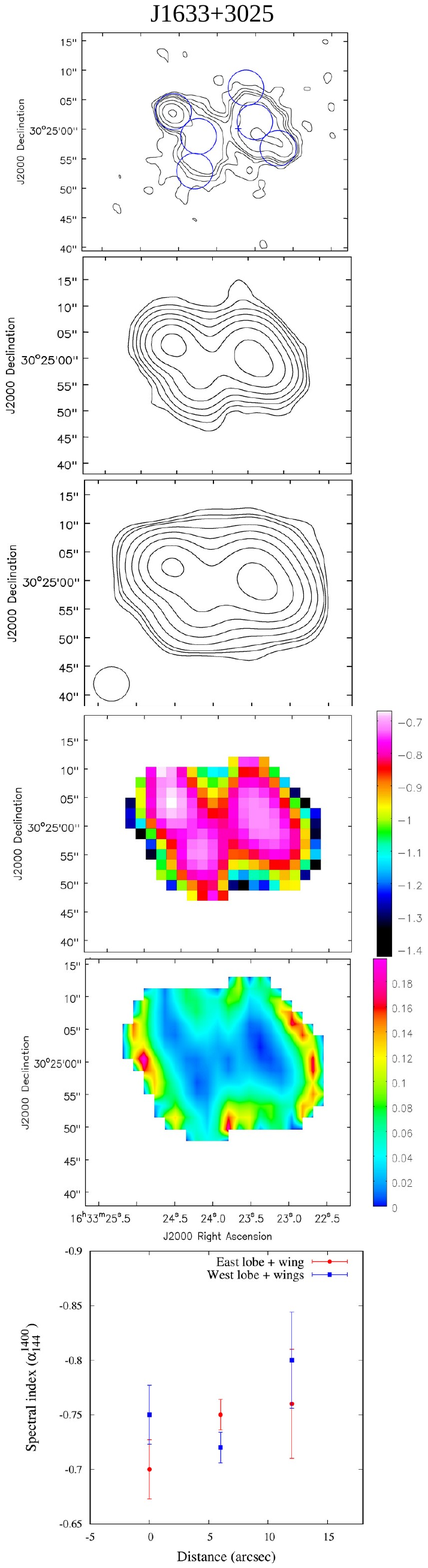}
\caption{Continued} 
\end{center}
\end{figure*}

\label{lastpage}
\end{document}